# Survey of Strong Authentication Approaches for Mobile Proximity and Remote Wallet Applications - Challenges and Evolution

Amal Saha
Email: amal.k.saha@gmail.com
Tata Institute of Fundamental Research (TIFR)
Mumbai, INDIA

Sugata Sanyal
Email: sugata.sanyal@tcs.com
Research Advisor
Tata Consultancy Services, Mumbai, INDIA

**ABSTRACT**

Wallet may be described as container application used for configuring, accessing and analysing data from underlying payment application(s). There are two dominant types of digital wallet applications, proximity wallet and remote wallet. In the payment industry, one often hears about authentication approach for proximity or remote wallets or the underlying payment applications separately, but there is no such approach, as per our knowledge, for combined wallet, the holder application. While Secure Element (SE) controlled by the mobile network operator (i.e., SIM card) may ensure strong authentication, it introduces strong dependencies among business partners in payments and hence is not getting fraction. Embedded SE in the form of trusted execution environment [3, 4, 5] or trusted computing [24] may address this issue in future. But such devices tend to be a bit expensive and are not abundant in the market. Meanwhile, for many years, context based authentication involving device fingerprinting and other contextual information for conditional multi-factor authentication, would prevail and would remain as the most dominant and strong authentication mechanism for mobile devices from various vendors in different capability and price ranges. EMVCo payment token standard published in 2014 tries to address security of wallet based payment in a general way. The authors believe that it is quite likely that EMVCo payment token implementations would evolve in course of time in such a way that token service providers would start insisting on device fingerprinting as strong means of authentication before issuing one-time-use payment token. This paper talks about challenges of existing authentication mechanisms used in payment and wallet applications, and their evolution.

## 1. INTRODUCTION

Wallet application has been variously described and there are two dominant forms related to payment, namely, proximity wallet and remote or cloud wallet [26]. Proximity wallet consists of rich user interface and a mobile application programming interface based communication with a secure payment application often running in Secure Element outside the mobile operating system or equivalent software container running in the operating system of mobile device and payment cryptogram created in this way may be exchanged with NFC-enabled point-of-sales (POS) terminal. Recent introduction of Host Card Emulation (HCE) [16, 17, 18] and EMVCo payment token [13] took a step forward, by introducing concept of one-time-use payment token for proximity payment. The latter may be obtained from the issuer or the card network like Visa or MasterCard, after authentication. These payment tokens would act as virtual or emulated payment card in card emulation mode of NFC and EMV payment cryptogram may be generated in application running in mobile operating system rather than Secure Element and may be exchanged with NFC enabled POS terminal.

Wallet may be described as a container or holder application used for configuring, accessing and analysing data from underlying payment applications, e.g., proximity payment or remote payment. In case the underlying payment application supports card, as in proximity payment, it conforms to EMV contactless specification [27] while the user interface of the container wallet application communicates with the payment application conforming to specifications published by Global Platform [2, 3]. For in-store payment, the authentication of the secure payment application, initiated by a POS terminal, in a chip based contact card is well defined and EMV specifications cover online and offline authentication processes. For EMV payment, the online authentication is based on symmetric key encryption while offline authentication process is based on public key cryptography.

However, there is no standardisation of authentication to the proximity wallet application which acts as a container of the actual payment application and other features like transaction history, etc.

In order to authenticate users and grant access to the Secure Element, first generation Google Wallet [26] for smartphone required a 4-digitnumeric PIN to launch the wallet application for the first time. Traditional physical credit cards do not have this feature and this had been claimed to be a differentiating factor.It may be noted that the first-generation Google Wallet stored the hash of 4-digit PIN and associated salt outside the SE. In fact the PIN and salt were stored in protected space on the file system of the mobile device, i.e., a space directly controlled by the mobile operating system [1]. This was reported to be vulnerable by security researcher [1] and Google changed the design by moving them to SE. This however implied taking approval of the manufacturer and the mobile network operator who control the space in SE thru cryptographic means and change of ownership of securing the PIN to the service providers like bank, transit companies, etc. This multi-party solution is however is not good from business perspective because everybody wants a pie of the business and hence there is conflict. HCE [16, 18] was aimed at addressing this problem.

Remote wallet refers to the scenario where payment details are stored on the cloud (i.e., server) and not locally on the mobile device. For authentication from mobile device to remote wallet system and also for authenticating remote payment transaction, soft token which generates TOTP algorithm based OTP locally on the mobile device or PC, but time-synched with the OTP server hosted in data centre of the wallet provider, is often used. Soft token runs on mobile device or the PC or the tablet. OTP features may be configured on the server and an instance of OTP would typically remain valid for a limited period of 30 to 60 seconds. One can have 6-digit OTP instead of 4-digit one. Another choice is hard token that generates TOTP algorithm based OTP. Hard token would be more secure because it can truly make the OTP out-of-band.

The problem with OTP based solution approach is that one needs to configure soft token as part of wallet application installation. But this is already in use as second factor of authentication, password or question/answer or device ID or MSISDN (Mobile Station International Subscriber Directory Number or mobile number) often being the first factor. Therefore this is no longer a challenge except impacting user experience to certain extent. Again, there is no standardisation for authentication to remote wallet.

The proximity wallet should work even when there is no connectivity of the mobile device with the server/cloud and an offline authentication is actually desirable. This does not however preclude use of one-time online provisioning of personalisation and sensitive authentication data needed from the server to the device. In case of HCE, some one-time-use payment tokens are obtained from the server after authentication and one-time-use, but valid-for-limited-period, payment tokens are retrieved in advance. Remote wallet, on the other hand, requires the mobile device or the PC to be connected with the cloud thru the internet and also the server to authenticate the client of the wallet and therefore an online authentication is more desirable. Authentication mechanism of the two wallet types is therefore different.

In the payment industry, one often hears about authentication approach for proximity or remote wallets separately, but there is no such approach, as per our knowledge, for combined wallet.

Purpose of this paper is to elaborate strong authentication process for wallet system which supports both proximity and remote payment features and appears as seamless authentication to the end user.

### 1.1 Review of Authentication Approaches in Mobile Payment Applications — Challenges and Evolution

Classic two-factor authentication based on hardware token has been existent for many years [20] and has been successful in certain domain like enterprise application and banking service. The token used here is a physically separate device [7] and token to be used as second factor of authentication and generated in this way, cannot be intercepted by malware like key-logger.

Both IBM X-Force research report [9] and Trustwave research report [10] highlighted malware attack on mobile devices, particularly, Android devices. According to Trustwave [10], the Android platform continues to be the focus of malware and in 2012, malware for Android grew 400%, from 50,000 to over 200,000 samples.

Advanced persistent threat (APT) has been reported by security researchers and one important means of stealing sensitive information by hackers and criminals thru APT is to install sophisticated malware on an end-point like mobile device or laptop or PPC. Multi-factor authentication is a way of defeating such attack.

One-time-password (OTP) has been proposed as a measure to defeat the malware attack. SMS-based out-of-band OTP for browser-based internet banking is effective and is widely used in online banking transaction originated on laptop or desktop machine. The user however has to key in the OTP sent thru the SMS and user experience suffers. In mobile environment, the mobile device itself may act as the OTP receiving device. Often one has a scenario where the user is running the business application, e.g., payment application, on the same mobile device which is also the OTP receiving device (OTP is generated on the server side) and security associated with the authentication of the user is considerably weakened in such case. On-device OTP is not-of-band, may be intercepted by malware and is therefore less secure, although user experience is better.

As a next candidate, public key infrastructure (PKI) may also be used for authenticating the user in a two-factor authentication scheme where user ID and password or PIN would act as first factor and the user's certificate as second factor where mutual authentication based on PKI is to be enforced. No doubt, PKI is a high standard for authentication of the user. It too has Achilles' heel -- the user private key associated with the digital certificate may not be stored securely, in all configurations and on all devices. Storing authentication credentials on Secure Element (SE) [25] is secure, but introduces dependency among mobile network operator and the device manufacturer.

Secure Element (SE) could secure OTP and PKI based user authentication considerably provided user interaction with the SE is protected from malware attack. Although Smartcard or SE provides processing and secure storage capabilities, it has no direct access to user interface (data input by user and display of data) of the parent device and this would make additional demand on security features provided by the device. Smartcard or SE based solution would therefore be vulnerable to malware interception attack. When the user is allowed to download apps from app-stores and other sources and install them on mobile device, at least one additional control is needed to secure it from malware attack. In a relatively controlled environment of Point-of-Sale (POS) terminal, the situation is relatively good, but not so in consumer devices like smartphones. Software based OTP or PKI based authentication, without using hardware SE, have been reported by many vendors, but it is vulnerable to malware attack.

Furthermore, the trend in hardware technology is moving towards building entire solution including processor, I/O, memory, cryptography, etc on a single chip. Problem of using smartcard or SE as separate from the chip is driven by commercial considerations. It is technically feasible to overcome the security issues associated with smartcard data input and display, mentioned above, by building additional smartcard-like security controls in the mobile device itself. Relevant question is— should one allow device hardware to support external Secure Element (hardware component outside the single chip) or should one build SE capabilities in the integrated hardware itself?

There has been investigation [5] to compare strengths of authentication in Trusted Execution Environment (TEE) with classic out-of-band two-factor authentication. TEE in this comparison is characterised by [5, 11] the following features:

- isolated execution to ensure applications execute completely isolated from
- unhindered by others and guarantees that code and data are protected at run-time
- secure storage to protect persistently stored data (e.g., cryptographic keys) belonging to a certain application from being accessed by other applications
- remote attestation to enable remote parties to ascertain they are dealing with a particular trusted application on a particular TEE
- secure provisioning to enable communication by remote parties with a specific application on a specific TEE while protecting integrity and confidentiality
- trusted path which is a channel for the user to input data to the TEE and for the TEE to output data to the user; the channel protects against eavesdropping and tampering.

There have been two major approaches to TEE [5] –one from Global Platform (GP) and the other from Intel [29]. The GP approach is defined thru GP specifications and work by Trustonic and semiconductor intellectual properties (IPs) from ARM [4, 8]. The second known TEE approach [6] is Intel's Identity Protection Technology (IPT). Samsung KNOX [6] approach for Android device is aligned with ARM specifications supporting GP TEE. Samsung KNOX has a container, a virtual Android environment within the mobile device, completed with its own home screen, launcher, applications, and widgets. ARM TrustZone®technology is a system-wide approach to security on high performance computing platforms. This hardware architecture combined with TEE software forms the basis of trust for a wide array of applications. New versions of Windows Phone and Apple iPhone devices too support trusted execution environment of some sort.

The number of mobile devices complying with TEE of any sort (GP or Intel) is limited [4]. Gemalto, ARM and G&D are driving efforts in this regard thru Trusttonic joint venture. One needs a solution that is not dependent on device manufacturer and is accepted widely in the industry. Push from Trustonic is trying to increase adoption of TEE in mobile device and it is being aligned with Trusted Service Manager (TSM) [28] needed for proximity payment. It may be pointed out that there are three known from factors of SE, namely UICC/SIM from mobile network operator, embedded SE from mobile device manufacturer and detachable micro-SD card from service provided. TEE solution may work without any secure element (UICC/SIM) from mobile network operator (MNO) because a hardware component needed to support TEE solution may be in the form of hardware component in the integrated chip. Since management of security of the embedded SE in TEE approach would require involvement of TSM and also ARM has significant influence on manufacturers of semiconductor integrated chip used in mobile devices, this approach is could gain momentum in future.

Currently Intel's share of integrated chip for mobile devices is limited. GP TEE approach is flexible and Intel's TEE solution may support GP TEE specification in future and even Intel's own TEE solution for integrated chip may get more traction and this may lead to availability of TEE enabled mobile device and containment of malware on mobile devices, particularly Android devices which have the largest market share.

Trusted Computing or TEE based devices could be a long-term approach towards addressing the authentication challenges. It appears that such devices would be in abundance by 2017.

It may be highlighted that EMVCo payment tokenisation [13] insists on authentication mechanism built into the process of issuance of payment token which acts as one-time-use virtual card. This however depends on existing authentication standards and token service provider has the option of choosing the right one. For cloud-based transaction, EMVCo payment tokenisation by way of creating virtual card is a good technique to secure it. As for solutions that implement host-based card emulation (HCE) [18], they may leverage tokenisation in different ways. It is not just about replacing the real card by a token; it is also about secure authentication of the token requesting (mobile) device and the cardholder using it. In this standard, token service provider may use simple to very strong authentication mechanisms including device fingerprinting [12, 15, 19, 21, 22, 23]. Google did exactly this in the latest implementation of Google Wallet [26]. Somewhat similar approach of reputation checks [30, 31] have been reported in intrusion detection systems.

In October, 2014, with iPhone 6, Apple introduced secure NFC payment [14]. With Apple Pay, during registration, user adds actual credit and debit card numbers to backend called Passbook and a unique Device Account Number is assigned, encrypted, and securely stored in the Secure Element, a dedicated chip in iPhone. These numbers are not stored on Apple servers. And when the user makes a purchase, the Device Account Number, along with a transaction-specific dynamic, one-time-use security code called token, is used to process payment. Actual credit or debit card numbers are never shared by Apple with merchants or acquirers, or transmitted over the network. Even with Apple's NFC payment, for retrieval of EMVCo-like payment token, authentication for issuance of token is involved.

Meanwhile, for many years, context based authentication (CBA) involving device fingerprinting [15, 19, 21, 22, 23] and other contextual information for conditional multi-factor authentication, would prevail and would remain as the most dominant and strong authentication mechanism for mobile devices from various vendors in different capability and price ranges. In fact, the authors believe that it is quite possible that HCE would evolve to insist on device fingerprinting as part of authentication before issuing payment token.

The above mentioned authentication approaches are general and now one will focus on applying them to the wallet application, i.e., the container application.

## 2. PROPOSED STRONG AUTHENTICATION APPROACHES FOR WALLET

### 2.1 Proposed Strong Authentication Approach for Proximity Wallet

Global Platform [2, 3] has defined specifications for smartcards, devices and backend processing that enable several parties to independently and securely manage their role in a single Secure Element. Applications from different stakeholders will rely on backend, device and on Secure Element (SE). An API referred to as the Secure Element access API, is used by the device applications to exchange data with their counterpart applications running in the Secure Element. Restricting the use of such an API is necessary since modern mobile operating systems do not efficiently prevent unauthorised access. Secure Element access control mechanism may be summarised as use of key and-or certificate by an owner of the SE to create Security Domain (SD) and then setting controls on the applications which may run in the SD. The owner of an SE and also the SDs may in turn allow other stakeholder to create a new SD and protect it with keys and-or certificates, similar to modern hotel room renting approach.

Using GP provisioning approach, one may develop an application that will validate the wallet (container application) authentication PIN (numeric values), separate from the authentication PIN used by proximity payment application conforming to EMV contactless payment specifications [27] and the former will run in an SD in Secure Element in the mobile device. While this approach will survive attack of jail-breaking or rooting of the mobile device, this is still vulnerable to malware attack.

Once the one-time provisioning thru back-office of Trusted Service Manager (TSM) is completed, connectivity of the device to the backend is not needed for offline authentication where the credentials would be stored securely on the SE and access control would be defined appropriately.

Trusted User Interface [3], part of GlobalPlatform specification Trusted Execution Environment (TEE) specification can prevent key-logging attack. In absence of TEE support, key-logging malware attack on SE access in the context of wallet authentication based on PIN, may be defeated if a virtual numeric keypad with shuffling feature, controlled by the application logic, is used.

Given the fact that TEE adoption in the mobile devices is insignificant as of now, this is a cheap and viable solution until TEE is widely supported in mobile devices in future.

User may enter the PIN using application controlled secure key pad, and initiate the end-to-end provisioning process between the device resident SE and the backend. The provisioning process would manage access control and encryption key.

For the scenario where SIM is the SE, mobile network operator (MNO) would be an important stakeholder and MNO's SD may be the starting SD for creating the secondary SD needed to host the wallet authentication application. While this would lock the proximity Wallet to the MNO, this is perhaps the most common scenario in Europe and hence commercially viable.

For the above solution to work, MNO may own the "Wallet PIN Authentication" application to be installed in SE and its lifecycle. Deployment of the application may be done in MNO backend before delivery of the UICC to the user or provisioned OTA.

If the HCE [18] based NFC payment is used by the wallet, relevant authentication would be dictated by EMVCo tokenisation standard compliant tokenisation service provider [13].

## 2.2 Proposed Strong Authentication Approach for Remote Wallet

Remote Wallet authentication has to use validation of credentials on the host side, i.e., ONLINE validation is needed. One has the option of having multi-factor authentication.

Mobile device ID (IMEI) would be sent by the application and there would validation on the host. Soft OTP token with generates known algorithm based OTP locally (TOTP algorithm is preferred), but time-synched with the OTP server hosted in data centre of the wallet provider, may be used as second factor of authentication.

If OTP soft token runs on hardware or separate mobile device, second factor OTP based authentication is still out-of-band and is very secure. But this is not user friendly.

OTP Soft token SDK may be used for embedding it in Wallet App for convenience (usability) so that the user does not have to key in the OTP. But static PIN needs to be entered by the user to unlock the token. Again application controlled keypad with shuffling feature may be used to reduce the attack by key-logging malware.

It must be noted that static PIN used to unlock the soft token is different from dynamic OTP PIN.

OTP token may be installed on PC as well and the same technique may be used on PC.

The system will provide interfaces to erase token data in case of pre-defined number of incorrect PIN attacks using the token policy setting of the server-side identity and access management system that will generate OTP. Immediate revocation of existing OTP token and dynamic re-seeding of OTP soft token are also typically supported.

The weakness of OTP soft token based solution approach is that you need a configured soft token as part of provisioning at the time of wallet installation. The most important challenge is that the static enabling PIN of the OTP token of the keying in of the generated OTP onto the application input screen (mobile handset app or browser on PC) is vulnerable to malware attack in case of in-band OTP soft token. Even if the OTP token is in SE instead of the rich operating environment of the mobile device, this vulnerability exists for in-band OTP soft token and mechanisms like secure input and display are needed as additional controls as already explained in section 1.1. TEE may help to implement these controls in case of in-band OPT soft token.

Even if PKI instead of OTP token, it is vulnerable to malware if user private key and the certificates are stored on the device and not in the SE. If private keys and certificate are stored in SE and cryptographic operations take place in SE of a mobile device on which remote payment transaction is initiated, an enabling PIN would be needed (typically a password based encryption scheme is use to protect the key store and the PIN/passphrase is needed) and the usual vulnerability associated with PIN entry and display would be present in mobile device because of in-band nature of the second factor authentication, without TEE support. In the absence of TEE support, in the interim period, application controlled virtual shuffling PIN pad would help to a certain extent, to secure the process.

Device fingerprinting and context based authentication (CBA) may be used as strong authentication for remote wallet in absence of Trusted Computing and is recommended and would prevail in coming years.

## 2.3 Proposed Approach of Strong Handset User Authentication for Combined Wallet

For a mobile handset user having access to combined wallet application installed on his/her mobile handset, secure element based authentication, with application logic controlled secure keypad and secure display, for connected device is recommended, when

TEE or Trusted Computing is not supported by the mobile device hardware. Remote payment using remote wallet requires the mobile device to be connected. Proximity payment, on the other hand, does not require the mobile device to be connected. In case of disconnected mobile device, authentication of wallet may limit the scope of the authorisation or entitlement to proximity wallet features only, a subset of features of the combined wallet and in case of connected mobile device; authentication may not put such limitation on authorisation or entitlement. For actual implementation of strong authentication in combined wallet application, the approaches discussed in previous sections may be chosen.

## 3. APPLICATION OF THE APPROACHES AND EVOLUTION

The previous section already highlighted usage scenarios of the proposed wallet authentication approaches, along with challenges like adoption of the mobile devices equipped with certain advanced technologies. Both long-term and medium-term strong authentication adoption have been highlighted.

A perspective on evolution of strong authentication for wallet has been provided.

## 4. CONCLUSIONS

Wallet may be described as container application used for configuring, accessing and analysing data from underlying payment applications. In the payment industry, one often hears about authentication approach for proximity or remote wallets separately and it is not common to see a combined treatment of the issue across wallet types. This paper throws some light on authentication and authorisation of a combined wallet. Secure element based authentication, with application logic controlled secure keypad and secure display, for connected device is recommended, when Trusted Execution Environment (TEE) and Trusted Computing is not supported by the mobile device hardware. Remote payment wallet requires the mobile device to be connected to backend service. Proximity payment, on the other hand, does not require the mobile device to be connected to backend service. In case of disconnected mobile device, wallet authentication may limit the scope of the authorisation or entitlement to proximity wallet features only, a subset of features of the combined wallet and in case of connected mobile device; authentication may not put such limitation on authorisation or entitlement.

Suggested longer-term approach of the wallet authentication is to leverage TEE or Trusted Computing in supported mobile device where a hardware component in integrated chip would act as embedded SE and it appears such devices would be in abundance by 2017. This is a win-win scenario for different stakeholders in the payment and related industries, namely mobile network operators, trusted service providers, semiconductor device makers and service providers like banks, transit companies, etc.

Meanwhile, for many years, context based authentication (CBA) involving device fingerprinting and other contextual information for conditional multi-factor authentication, would prevail and would remain as the most dominant and strong authentication mechanism for mobile devices from various vendors in different capability and price ranges.

EMVCo payment token standard has been introduced in 2014 and new and tries to address authentication and other security issues in the context of payment. The authors believe that it is quite likely that EMVCo payment tokenisation implementations would evolve in such a way that token service providers would start insisting on device fingerprinting as strong means of authentication before issuing one-time-use payment token.

This paper talked about challenges of existing authentication mechanisms used in payment and wallet applications, and their evolutionary path. In future, adoption of new approaches highlighted in the paper would get momentum in mobile payments.